# Taverna Mobile: Taverna workflows on Android

Hyde Zhang, Stian Soiland-Reyes✉, Carole Goble

University of Manchester, United Kingdom

### Abstract
Researchers are often on the move, say at conferences or projects meetings, and as workflows are becoming ubiquitous in the scientific process, having access to workflows from a mobile device would be a significant advantage. We therefore have developed Taverna Mobile, an application for Android phones which allows browsing of existing workflows, executing them, and reviewing the results.

## Motivation and Objectives

We present Taverna Mobile, a mobile application for browsing and executing Taverna Workflows on Android phones. Taverna is a mature scientific workflow tool suite, designed to combine distributed web services and local tools into complex analysis pipelines (Wolstencroft *et al.*, 2013). Taverna can be executed on local desktop machines, using the Taverna Workbench, or larger infrastructures such as grid or cloud installations using the Taverna Server. Although Taverna has been growing in popularity within fields such as astronomy, chemistry, biodiversity and text mining, its largest user base remains within the bioinformatics community.

The social networking site myExperiment facilitates sharing and reuse of scientific workflows within the scientific community (Goble *et al.*, 2010) and currently includes more than 1,000 Taverna workflows[1], most of which are freely accessible under a Creative Commons license.

Researchers are often on the move, say at conferences or projects meetings, and as work-flows are becoming ubiquitous in the scientific process, having access to workflows from a mobile device would be a significant advantage. We therefore have developed Taverna Mobile, an application for Android phones which allows browsing of existing workflows, executing them, and reviewing the results.

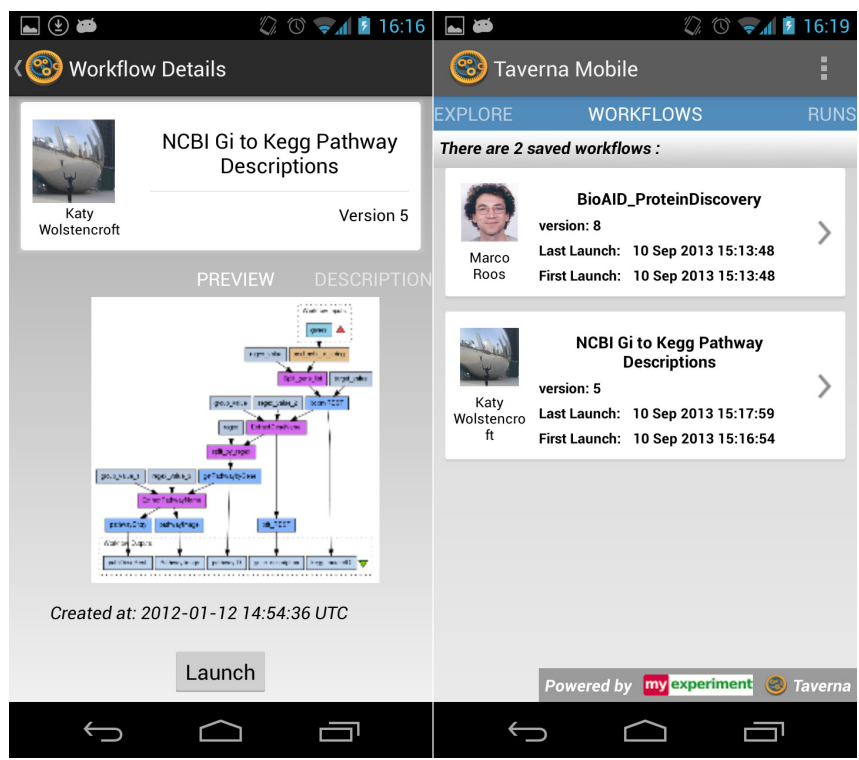

Figure 1. Left: Taverna Mobile browsing myExperiment workflow #2659 NCBI Gi to Kegg Pathway Descriptions. Right: Previously run workflows are saved on the device, with their earlier inputs.

---

1    http://www.myexperiment.org/workflows

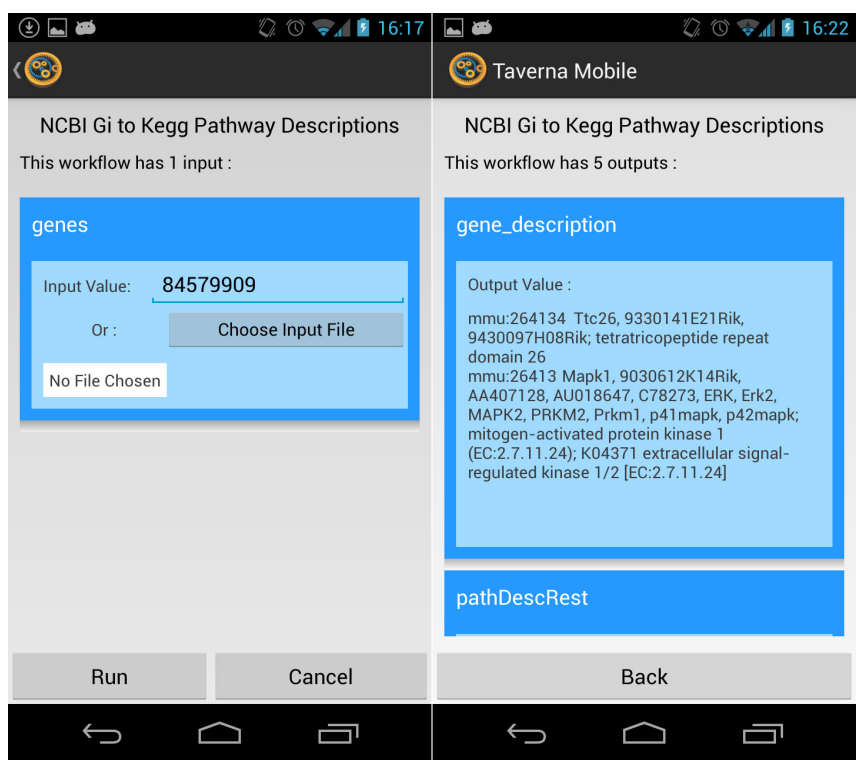

Figure 2. Left: Taverna Mobile prompting for inputs before launching workflow #2659, filled with Gene Identifier (GI) 84579909. Right: Results from workflow run, showing KEGG Pathways with descriptions.

## Methods

The mobile user can browse and search existing workflows from myExperiment. Metadata about workflows, including uploader, description and rating are displayed together with a diagram of the workflow. From here the user can either "favourite" a workflow to explore it in detail later, download the workflow definition, or click "Run" to execute the workflow.

For running workflows, Taverna Mobile connects to a remote installation of the Taverna Server. The user is prompted for workflow inputs; which can either be typed in or selected from a file on the device. The Android application initiates workflow execution on the server, which status can be monitored from the mobile. Finally, workflow outputs are retrieved and stored as files on the Android device, from where they can be viewed, saved to external locations such as Dropbox and Google Drive, or shared by email.

Frequently accessed workflows are remembered by Taverna Mobile, including their previous input data, which allow the user to quickly repeat workflow runs on the go. Long-running workflow runs that have been started on the Taverna Server through other means, such as through a portal, can also be monitored in the application.

See Figures 1 and 2 for detailed examples making reference to myExperiment workflow #2659 NCBI Gi to Kegg Pathway Descriptions[2] and #74 BioAID Protein Discovery[3].

## Results and Discussion

Taverna Mobile does not aim to reproduce the full experience of building workflows in the Taverna Workbench, rather it focuses on tasks we have deemed relevant to a scientist that is not at her desk. For instance, when visiting a conference she might hear about someone's workflow, which she can quickly locate and mark for later exploration. When in the biology lab, faced with updated scientific data, the scientist can rerun her own workflow with new inputs. While commuting, she can monitor the status of a long-running job.

---

2  http://www.myexperiment.org/workflows/2659/versions/5
3  http://www.myexperiment.org/workflows/74/versions/4

Taverna Mobile support researchers in these situations by providing immediate access to workflows and workflow runs. The Taverna Mobile application is planned to be released on the Google Play market by the end of 2013, and its source code is available on the Taverna-Mobile[4] project.

## Acknowledgements

Hyde Zhang developed the Taverna Mobile application as part of his BSc third year project and a summer placement at School of Computer Science, University of Manchester, and was supervised by Professor Carole Goble. He would also like to thank Robert Haines, Alan Williams and Donal K. Fellows from the myGrid team for their great technical support.

Stian Soiland-Reyes wrote this abstract together with Hyde Zhang, and is funded for the Wf4Ever project by the European Commission's 7th FWP FP7-ICT-2007-6 270192.

---

4    https://github.com/myGrid/taverna-mobile